\documentclass[prb]{article}

\usepackage{chemformula} 
\usepackage[T1]{fontenc} 
\usepackage{graphicx}
\usepackage{dcolumn}
\usepackage{bm}
\usepackage[utf8]{inputenc}
\usepackage{mathptmx}
\usepackage{etoolbox}
\usepackage{booktabs}

\author{Wagner Gomes Rodrigues Junior\\
Institute of Exact Sciences and Technology, Federal University of Amazonas\\
wagnerif@ufam.edu.br\\
V. B. Henriques\\
Physics Departament, S\~ao Paulo University\\
 verahenriques@usp.br}

\title {Characterization of porous nanoparticles using the lattice Boltzmann method for fluid flow.}

\begin{document}

\maketitle

\begin{abstract}
Nanoporous capsules have been the subject of intense investigation in the field of drug delivery. One of the essential properties of such particles, which requires characterization, is their structure. Many experimental techniques have been used for this purpose, such as wide-angle neutron or X-ray scattering, or light scattering. Herein, we report theoretical data on the relationship between the size, porosity and drag force of porous particles. Data were obtained from a numerical procedure based on the lattice Boltzmann method for the calculation of the creeping flow over porous spherical objects. Previous analytical solutions to the hydrodynamic equations yielded different predictions for the drag force under conditions of high permeability. In the light of our findings, we compared our data to previous divergent analytical solutions and analyzed experimental results for dynamic and static light scattering of porous membrane vesicles Our findings strongly indicate a potential source of error that has led to inconsistencies in DLS measurements when compared to other techniques.
\end{abstract}

\section{Introduction}

Viscous flow through porous media is a subject of interest in various applied areas, from geological to biological science \cite{bear2012-livro, vafai2015handbook}. Porous shells are used as models for the study of microcapsules in the field of drug transport \cite{baeza2017recent, Sayed2017-porous-particle-review}, or as an experimental model, in the study of biomembranes \cite{heimburg-book}. Their interaction with fluids constitutes an area of intensive study.

In particular, nanoporous capsules designed for delivery \cite{baeza2017recent} require investigation of their structure and flow properties.  Wide-angle neutron (WANS) and small-angle X-ray (SAXS) scattering, transmission electronic microscopy (TEM), as well as light scattering are some of the techniques that have been used for the characterization of hollow particles \cite{chen2012-pore-structure-various-techniques, kaasalainen2017-structure-lightscatt}. In the case of dynamic light scattering (DLS), data on diffusion are used to obtain the particle radius using the Stokes Einstein equation for the relationship between drag force and diffusion, which is valid for solid particles. In the case of porous particles, this relationship must be redesigned, and the investigation of this question is the aim of the present study.

On the theoretical side, the analytical study of creeping flow through porous objects implies dealing with the choice of the equation that describes the flow inside the porous system, as well as with the boundary conditions at the interface between the porous object and the free fluid flow. For a low Reynolds number, the flow inside the porous object has been described, either in terms of Darcy’s equation \cite{jones1973low}, under different boundary conditions, or in terms of a Navier-Stokes equation with a Darcy term added \cite{neale1974creeping}, which has been called the Brinkman approach. Different analytical, numerical and experimental studies in the last two or three decades have led most of the interested community to accept the second alternative as the correct approach \cite{jaiswal2015stokes}.

However, analytical treatment of hydrodynamic flow through porous materials is not always possible, 
and computational fluid dynamics, in which the hydrodynamic equations are discretized, are frequently used in the case of complex boundary conditions. An alternative numerical technique for the investigation of fluid dynamics is the lattice Boltzmann method (LBM): instead of looking directly at the hydrodynamic equations, one may resort to kinetic theory for the evolution of the particle velocity distribution, in a discretized version of the Boltzmann equation. The relationship between the molecular Boltzmann approach and the macroscopic hydrodynamic description is established by imposing conservation of mass and momentum \cite{kerson1963statistical}. A discrete version of the Navier-Stokes equation can be obtained from the discretized Boltzmann equation in the linear approximation for the collisional integral \cite{Chen, Higuera}.

The LBM is a powerful technique for computational modelling in the study of complex fluids and has attracted the interest of researchers in computational physics in recent decades as an interesting alternative for the treatment of the interaction of different systems and fluids \cite{guo2002LB, monteferrante2014lattice}. Among the advantages of the LBM are its easy parallelization, the fact that it is not necessary to solve the Poisson equation for the pressure, its relatively simple implementation due to discretization of time and space, and specially the possibility of working with complex boundary conditions. This method has been successfully applied in multiphase microfluidics, porous media or charged colloidal suspensions \cite{huang2015multiphase, zhang2011lattice, yoshida2014transmission, horbach}.

In the lattice Boltzmann method \cite{Wolf}, speeds are described using a discrete set $v_i$ , and the discrete velocity distribution functions ($f_i$) evolve in time through the lattice of fluid cells. The hydrodynamic quantities are obtained as averages over the distribution functions. In this description, the boundary conditions are applied to the particular $f_i$s, and not to velocities and other hydrodynamic variables. For the Dirichlet boundary condition, the most common treatment is known as “bounce-back” \cite{ziegler1993boundary, he},  which produces results with first-order precision. Treatments for boundary conditions with more accurate results use interpolation techniques \cite{mei1999accurate, filippova1998grid}. However, Inamuro et al.\cite{inamuro1995non} showed that the error caused by the bounce-back is small if the relaxation time is close to 0.5.

Our study focuses on the flow past porous objects, which we undertake with the LBM approach in the single relaxation time version. Besides its success in the study of different complex fluids, it has been shown that the approach we applied gives \cite{prestininzi2015reassessing} accurate results in porous media studies for small Knudsen numbers ($Kn < 10^2$).

We investigated the drag force on porous spherical shells and spheres, under creeping flow (low Reynolds numbers), as a function of porosity and Reynolds number. We adopted  “partial reflecting” evolution conditions \cite{walsh2009new} for the distribution functions in the porous cells. As predicted in several studies, the drag force on the spherical objects is reduced as a function of permeability but suffers little impact from the Reynolds number for Reynolds numbers in the range $ Re <1$.

Our results for the porous spheres are compared to the analytical predictions of Jones \cite{jones1973low} and of Neale and colleagues \cite{neale1974creeping}, and to the experimental results of Masliyah et al. \cite{masliyah1980terminal}. These three studies agree in their predictions for the drag force in the case of low permeabilities. For larger permeabilities, for which Masliyah has no experimental results, the two analytical predictions diverge, with Neale’s result decaying more rapidly with permeability than Jones’s result. Jones uses Darcy’s equation for the porous region, with continuity of pressure and radial velocity components, and discontinuity of the tangential velocity component proportional to strain, in an adapted version of Beavers . and Joseph \cite{Beavers}condition for planar surfaces. On the other hand, Neale and collaborators use Brinkman’s equation for porous media, with continuity of the tangential velocity at the interface. In our study of the same problem, we obtained the drag force from the evolution of the velocity distribution functions, with partial reflection in the porous cells, and with a “reflection coefficient” dependent on porosity. The choice of boundary conditions on porous and fluid cells is independent of the macroscopic description. Surprisingly, our results reproduced the analytical predictions of Jones, based on the Darcy equation. This is an interesting result since it relates boundary conditions to hydrodynamic quantities and boundary conditions to the velocity distribution functions.

As an application, we tested our proposal against a specific problem in the study of particles of biological interest. Lipid vesicles present different thermal phases, which are related to the order of the hydrocarbon chains or lipid headgroups and gel-fluid phase coexistence, and these are clearly established for lipids with polar head-groups \cite{heimburg-book}. However, in the case of dissociating head-groups, special physical properties arise in the transition region. These can be explored through a full range of techniques, which range from thermal to spectroscopic or transport techniques \cite{lamy2003peculiar}. In particular, dissociation \cite{Barroso} accompanied by pore formation has been reported for dimyristoyl phosphatidylglycerol (DMPG) \cite{Thais, riske2009extensive}. 
Investigation of the vesicle size led to inconsistencies between the results obtained via DLS (dynamic light scattering) and SLS (static light scattering).  For uncharged lipid solutions, both techniques yielded very similar data; whereas, for the charged vesicle, DLS yields an effective radius that is approximately three times smaller than the value obtained via SLS. What could be the origin of such a difference? In SLS, particle size is obtained directly from extrapolation of scattering at angle zero.  As for DLS, the technique produces data on diffusivity, and particle size is obtained from the Stokes-Einstein relation.

SLS measurements suggest an increase in the vesicle radius by a factor of three in the transition region. Taking the SLS result as the correct one, a possible explanation for the increase in vesicle radius, presented by Enoki et al. \cite{Thais}, would be the swelling of the lipid particles, with the appearance of pores. The appearance of pores in giant unilamellar vesicles was observed by Riske et al. \cite{riske2009extensive}. The hypothesis behind this explanation would be that the DLS would not be able to distinguish between an integral and a larger porous particle of the same mass. Our proposal allows us to verify this rationale for the discrepancy found in the study of Enoki et al. \cite{Thais}

Our paper is organized as follows: in Section 2, we briefly describe the LBM  for the calculation of the fluid drag force; in Section, 3 we define important quantities and equations for the description of flow through porous objects and review some previous analytical results, which are written in revised form in order to allow comparison; in Section 4, our approach to the LBM simulations is described and the results are presented and compared; in Section 5, we discuss the relationship between drag and diffusion for porous particles; Section 6 provides our conclusions.

\section{Fluid flow and molecular probability distribution function}

Fluid flow may have different mathematical descriptions. Newtonian fluids may be described through a differential equation for local \textit{fluid} velocity $\mathbf{u}(\mathbf{x}, t)$, which is known as the Navier-Stokes equation. Alternatively, fluid flow may be described in terms of average molecular properties obtained from a \textit{molecular} distribution function $F(\mathbf{x}, \mathbf{v}, t)$. The molecular mass distribution function $F(\mathbf{x}, \mathbf{v}, t)$ obeys a differential equation, which is called the Boltzmann equation.  

  The two descriptions may be connected by the application of mass, energy and momentum conservation laws to the \textit{molecular} description. In the presence of a pressure gradient, this yields the \textit{fluid} Navier-Stokes equation.

There are no analytical solutions, at the present time, for any of the two differential equations, except for in very special cases. Therefore, one has to resort to numerical calculations, which necessarily imply discretization.

\subsection{The continuous case: relation between macroscopic and microscopic descriptions }

The Boltzmann equation describes the evolution in time t of the fluid mass molecular distribution function $ F \equiv F (\mathbf{x}, \mathbf{v}, t) $ under the effect of binary molecular collisions. A linear approximation for the collisional integral, also known as BGK \cite{he1997analytic}, yields the first order Boltzmann equation for the mass distribution of particle velocities, which is written as

 \begin{equation}\label{bbgk}
  \frac{\partial F}{\partial t}+ \mathbf{v}\cdot\mathbf{\nabla}F=-\frac{1}{\tau}(F- F^{eq}),
 \end{equation}
 
 in the absence of an external force. Here, $\tau$ is a relaxation time associated with the average time interval between collisions and $F^{eq}$ is given by a local equilibrium Maxwell distribution function.

 Macroscopic quantities such as density $ \rho(\mathbf{x},t)$ and fluid velocity $\mathbf{u}(\mathbf{x}, t)$ may be calculated from the moments of the distribution function $ F(\mathbf{x}, \mathbf{v},t)$ . Conservation laws \cite {Cercignani1} yield the hydrodynamic equations for the fluid. In particular, conservation of momentum under a pressure gradient $\nabla p$ leads to the Navier-Stokes equation for the local fluid velocity $\mathbf{u}(\mathbf{x}, t)$, given by

  \begin{equation}\label{navier}
  \frac{\partial \mathbf{u}}{\partial t } +  \mathbf{u}\cdot\nabla \mathbf{u} = -\frac{1}{\rho}\nabla p + \nu(\tau)\nabla^2 \mathbf{u},
  \end{equation}  
\noindent where $\rho$ is fluid density and $\nu$ is kinematic viscosity, related to viscosity $\mu$ through $\nu \equiv \mu / \rho$. Derivation of the fluid equation from the Boltzmann equation
allows one to connect the microscopic particle relaxation time $\tau$ to the macroscopic kinematic viscosity $\nu$ using:

\begin{equation}\label{nu-tau-analit}
\nu (\tau) =\frac{k_{B}T \tau}{m}
\end{equation}

\noindent where T is temperature and $m$ is molecular mass of the fluid particles. The relationship between macroscopic viscosity $\nu$ and microscopic relaxation time $\tau$ is made explicit by writing $\nu (\tau)$. 
 
\subsection{Discretized description}

The lattice Boltzmann equation (LBE) is a kinetic equation obtained from the Boltzmann equation, through discretization of velocity, space and time. The first step consists of representing fluid particle velocities $ \mathbf{v} $ through a discrete set of velocities $ \mathbf{v}_i $ so that the continuous distribution function $F(\mathbf{x}, \mathbf{v}, t)$ is represented through the corresponding set of distribution functions $f(\mathbf{x}, \mathbf{v_i}, t) \equiv f_i(\mathbf{x}, t)$, with $i$ indicating the specific velocity of the set $ \mathbf{v}_i $. 

The boundary conditions for the continuous Boltzmann equation (Eq. \ref{bbgk}) are described in the discretized version as forces on the solid boundaries of the fluid vessel.

\subsubsection{Discretized Boltzmann equation and conservation laws}

The discretization of the velocity space is not arbitrary, since isotropy and Galilean invariance must be preserved \cite{Wolf}. In this study, we use the $D3Q19$ model \cite{Qian}, for which $i = 0, . . . 18$ (see Figure 1). The chosen model guarantees isotropy and offers the best combination of accuracy and speed \cite{mei}.

 \begin{figure}[h]
\centering
\begin{quote}
\begin{center}
                \includegraphics[width=0.5\hsize]{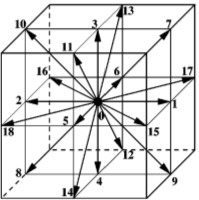} 
	   \caption{\textit{velocity vetors in the $D3Q19$}}\label{figura:D3Q19}               
\end{center}
\end{quote}
 \end{figure}

Equation \ref{bbgk} is then discretized in space and time and the new distribution functions evolve through a two-step process: streaming along lattice sites (propagation phase) and on-site collisions (collision phase). In the propagation phase, while in the collision phase, the $f_i$ are translated from each node to neighboring nodes, the $f_i$ relax to local equilibrium. After space and time discretization, and some manipulation, we may rewrite Eq \ref{bbgk} as a set of equations for the set of fluid particle distribution functions $f_i(\mathbf{x}, t)$:

 \begin{equation}\label{Boltz-evo-f}
   f_i(\mathbf{x}+\mathbf{c}_i\Delta t,t + \Delta t)- f_i(\mathbf{x},t)=-\frac{\Delta t}{\tau}(f_i- f_i^{eq}).
 \end{equation}

\noindent Here, velocities $\mathbf{c}_i$ , positions $\mathbf{x}$ and time intervals $\Delta t$ are written as dimensionless variables, with $c_i \equiv v_i/U$ , $\Delta x' \equiv \Delta x/L$ and $\Delta t' \equiv \frac{U}{L} \Delta t$ . $U$ is a reference fluid velocity and $L$ a typical linear dimension of the containing walls. 

 The local equilibrium distributions $f_i^{eq}$ for the $c_i$ are written as an expansion of Maxwell’s distribution for low local fluid velocities $\mathbf{u}$,

 \begin{equation}\label{Feq-Maxwell}
 f_i^{eq}(\rho,\mathbf{u})=W_i\rho\Big\lbrace 1+\dfrac{3U}{c_s^2}\mathbf{c_i}\cdot \mathbf{u}+\dfrac{3U^2}{2c_s^2}\left[\dfrac{3}{c_s^2}
 (\mathbf{c}_i\cdot \mathbf{u})^2-\mathbf{u}^2\right]\Big\rbrace. 
\end{equation}

\noindent where $c_s$ is root mean square molecular speed at a given temperature, with $c_s^2 = 3\frac{k_BT}{m}$, under the condition that fluid local speed $u$ must be much lower than $c_s$ \cite{Wolf} . Coefficients $W_i$ in Equation \ref{Feq-Maxwell} are fixed through the conservation equations. Mass density $\rho$, mass flux $j$ and energy $E$ must be preserved locally, which is guaranteed through the following equations:

\begin{equation}\label{densi}
 \rho(\mathbf{x},t)= m \rho \sum_i f_i(\mathbf{x},t),
\end{equation}

\begin{equation}\label{mom}
 \mathbf{j}(\mathbf{x},t)=\rho(\mathbf{x},t)\mathbf{v}(\mathbf{x},t)=mU\sum_i\mathbf{c}_i \rho f_i(\mathbf{x},t),
\end{equation}

\begin{equation}\label{Ener}
 E(\mathbf{x},t)=m U^2\sum_i\dfrac{|\mathbf{c}_i|^2}{2}f_i(\mathbf{x},t).
\end{equation}

To connect to macroscopic viscosity $\mu$, the rules governing collisions are chosen
such that the average movement of the fluid is consistent with the Navier-Stokes equation. In parallel to the continuous theory, coherence between the Boltzmann and Navier-Stokes equations in the discretized theory for the fluid also yields a link between the dimensionless relaxation time $\tau$ and the dimensionless kinematic viscosity\cite{Succi} $\bar{\nu} \equiv \frac{\nu}{U L} $ (see Eq. \ref{nu-tau-analit}). This is given by:

\begin{equation}
\bar{\nu} (\bar{\tau})=\frac{1}{3}\left(\bar{\tau}-\frac{1}{2}\right).
\end{equation}

 \subsubsection{The evaluation of force in the LBM}
 
 Fluid exerts a force on obstacles, which may be calculated from momentum exchange \cite{Ladd, Mei1}. The force in direction i on a solid cell at $\mathbf{x}_s$ due to fluid in a neighboring cell at $\mathbf{x}_s + \mathbf{c}_i \Delta t$ is given by:
 
\begin{equation}
 \mathcal{F}_{x_s, i}=-\frac{\Delta p_i(\mathbf{x}_s + \mathbf{c}_i \Delta t)}{\Delta t}
\end{equation}

Fluid in the adjacent cell at $\mathbf{x}_s + \mathbf{c}_i \Delta t$ suffers a change in momentum between $t$ and $t + \Delta t$, which is given by:

\begin{eqnarray}
 &&\frac{\Delta p_i(\mathbf{x}_s +c_i\Delta t)}{\Delta t} =\\
 &=& \left[f_{-i}(\mathbf{x}_s+\mathbf{c}_i \Delta t,t+\Delta t)(-c_{i})-f_i(\mathbf{x}_s +\mathbf{c}_i \Delta t,t +\Delta t)c_i\right]\frac{\Delta V}{\Delta t}.\nonumber
\end{eqnarray}
 
 \begin{equation}
 \frac{\Delta p_i}{\Delta t} = (f^{\prime}_{i}(\mathbf{x},t^{\prime})c^{\prime}_{i}-f_i(\mathbf{x},t)c_i)(\Delta V)
\end{equation}

where $f_{i}$ is the fluid probability density of the fluid particle with velocity in the $i$ direction after collision, and $\Delta V$ is the volume of one lattice node. It follows that the drag force $\mathcal{F}_d$ on the surface of a volume with border $C_s$ is given by:

\begin{equation}\label{FS}
 \mathcal{F}_d= \sum_{\mathbf{x}_s\in C_s}\sum_{i\neq 0}c_i(f_i(\mathbf{x}_s+\mathbf{c}_i\Delta t,t+\Delta t)+f_i(\mathbf{x}_s,t)).
\end{equation}
where $\mathbf{x}_s$ is a border node.

\section{Fluid flow through porous media}

  Important characteristics in the study of fluid flow through porous media are porosity $\phi$ and permeability $k$. Porosity $\phi$ is defined as the fraction of porous volume with respect to the total volume of the corresponding solid object. Alternatively, porosity may be parametrized through the equivalent solid fraction $n_s =1-\phi$. Permeability $k$ is defined as the ability of a given material to allow the passage of fluid. Darcy’s empirical law states that the volumetric flux q (rate of volume flow across a unit area) through the porous material decreases linearly with the pressure gradient, with the coefficient $k/\mu$ the ratio of permeability to viscosity:

\begin{equation}\label{darcy}
q = -\frac{k}{\nu \rho}\nabla P
\end{equation}
where $P$ is pressure. In more general cases, permeability is a second order tensor, but for isotropic media, permeability can be treated as a scalar.

A permeability dependent drag coefficient, $C_d (k)$, is defined as:

  \begin{equation}\label{Cd_k}
  C_d(k)=\dfrac{2\mathcal{F}_d(k)}{\rho U^2 A(R)},
 \end{equation}
\noindent in which $\mathcal{F}_d (k)$ is the drag force dependent on the permeability $k$ of the object in the fluid channel. $A$ is the object’s transverse section and depends on the radius $R$, in the case of spherical objects. U is the speed of the stationary fluid. For a solid sphere of radius R, the drag force is given by the Stokes relation (\cite{landau2013fluid}) $\mathcal{F}_d = 6\pi\rho\nu R U$ and the transverse section is $A = \pi R^2$. Thus, for an impermeable sphere, we have:

\begin{equation}\label{Cd_0}
C_d(k=0)=\frac{24}{Re}
\end{equation}
Additionally, a dimensionless drag coefficient may be defined, which is the ratio of the drag coefficients of porous and corresponding solid objects, and is given by:

\begin{equation} \label{r_k}
r(k)=\frac{C_d(k)}{C_d(k=0)} \equiv \frac{\mathcal{F}_d(k)}{\mathcal{F}_d(k=0)}.
\end{equation}
The drag coefficient is equivalent to the ratio between drag force $\mathcal{F}_d (k)$ on an object of finite permeability $k$ and the drag force on the corresponding solid object, $\mathcal{F}_d (k=0)$.

\subsection{Analytical predictions}
In analytical treatments, one must treat both the free fluid flow as well as flow through the porous object, which must obey adequate continuity conditions. 

Free fluid flow is governed by the Navier-Stokes equation (Eq. \ref{navier}). We are interested in situations of high viscosity, such that the viscous force dominates over the inertial forces\cite{kerson1963statistical}, characterized by a low Reynolds number $Re \equiv UL/\nu$. In this case, we may write a simpler Navier-Stokes equation (\ref{navier}):

 \begin{equation}
  -\frac{1}{\rho}\nabla p + \nu \nabla^2 \mathbf{u}=0.
  \label{navier-low-re}
  \end{equation}
Our interest lies in porous objects of spherical symmetry. Fluid flow through such objects has been investigated analytically by different authors \cite{jones1973low, neale1974creeping, qin1993creeping}. Free-fluid flow must naturally be described using Eq. \ref{navier-low-re}. However, the porous flow has been treated using distinct hydrodynamic equations that include permeability $k$ as a parameter. Special boundary conditions are imposed on the surface of the permeable solid, which also differ when using the two approaches.

  We present here the analytical results developed in those two previous studies \cite{jones1973low, neale1974creeping}, which predict coincident results at low permeability, but diverge as permeability increases. In order to compare these with the results of our simulation, and to make evident the difference between the two analytical predictions, we present the corresponding equations in a slightly modified form in relation to the original papers. Jones\cite{jones1973low} considers an equation for fluid velocity in the pores derived from Darcy’s phenomenological equation (\ref{darcy}), as proposed previously by Beavers and Joseph\cite{Beavers}.
This is\cite{Batchelor}:

\begin{equation}\label{darcy-velocity}
\frac{1}{\rho}\nabla P - \frac{\nu }{k} \mathbf{v} = 0.
\end{equation}

\noindent Note that we use $\mathbf{v}$ for fluid velocity inside pores, instead of $\mathbf{u}$, which is the free fluid velocity, in order to make the reader aware of the different fluid flow conditions. Moreover, the spherical symmetry requires that spherical coordinates are used, so that $ \mathbf{v}= \mathbf{v}(r, \theta)$ and $ \mathbf{v}= \mathbf{v}(r, \theta)$. As for boundary conditions, the author imposes continuity of pressure and conservation of mass (implying $u_r=v_r$) across the surfaces of the porous object, and a semi-empirical slipping condition for the component $r\theta$ of the rate of the stress tensor \cite{Beavers} on the surfaces, which are given by:

\begin{equation}
e_{r\theta}=\alpha(u_{\theta}-v_{\theta}.)
\end{equation}
\noindent $u_{\theta}$ is the velocity of the fluid on the surface of the porous object on the free fluid side, while $q_{\theta}$ denotes the velocity of the fluid inside the porous object.
The phenomenological parameter $\alpha$, related to the properties of the specific porous material, is introduced and should be obtained from an experiment.

Jones's \cite{jones1973low} calculations yield:

\begin{equation} 
r_{shell, J}(k)=\frac{\alpha + 2{\kappa}^{1/2}}{\alpha + 3 {\kappa}^{1/2} -6 {\kappa}^{3/2} +\frac{(3/2) \kappa (\alpha + 
6 {\kappa}^{1/2}) (3\gamma\kappa +
\frac{3 {\kappa}^{1/2}} {10\alpha}+
\frac{\gamma}{10})}{3\gamma\kappa+
(1-\gamma^3)(\frac{3}{10\alpha}{\kappa}^{1/2}+\frac{\gamma}{10})}}
\label{jones0}
\end{equation}

Here, $\kappa$ is the normalized permeability $\kappa\equiv \frac{k}{R^2}$, also known as the Darcy number. $R$ is the sphere radius, $\gamma$ is the ratio of internal to external radius of the shell, and $\alpha$ is a dimensionless constant that depends on the properties of the porous solid.

When the inner radius tends to zero, $\gamma\rightarrow 0$, we have a porous sphere. In this limit:

\begin{equation}
r_{sphere,J}(k)= \frac{\alpha + 2{\kappa}^{1/2}}{(\alpha + 3 \kappa^{-1/2} +\frac{3}{2}\alpha \kappa +3 \kappa^{3/2})}
\label{jones}
\end{equation}
When, $k\rightarrow 0$, which corresponds to the case of an impermeable sphere, $\beta \rightarrow \infty$, the Stokes drag force prevails, and

\begin{equation}
r(k\rightarrow \infty)=1,
\end{equation}
\noindent as expected.
Almost simultaneously, an alternative treatment is proposed by Neale and collaborators \cite{neale1974creeping} on the basis of the model proposed by Brinkmann \cite{brinkman1949calculation}. The authors\cite{neale1974creeping} argue that the 1st order Darcy differential equation (\ref{darcy-velocity}) for the porous region cannot give a precise description for the boundary layer in the permeable region. They present several results from previous treatments of the Darcy flow under different boundary conditions, but argue that flow inside the permeable object must obey the Navier-Stokes equation (Eq. \ref{navier}), with an additional Darcy permeablity term (see Eq. \ref{darcy-velocity}). Continuity of pressure of both velocity components and of the stress tensor $\tau_{r \theta}$ is then adopted for the interface between the fluid and the solid surface. The modified Navier-Stokes equation for the fluid velocity inside pores, $\mathbf{v}$,  is given by:

\begin{equation}
-\frac{1}{\rho}\nabla P + \widetilde {\nu} \nabla^2 \mathbf{v} - \frac{ \nu }{k} \mathbf{v} = 0,
\end{equation}

\noindent where $\widetilde {\nu}$ is any effective viscosity.

This second approach yields the following expression for the drag ratio on porous spheres \cite{neale1974creeping}:

\begin{equation}\label{Neale}
r_{sphere, N}(k)= \frac{2[1-\kappa^{-1/2}tanh(\kappa^{1/2})]}{2+3\kappa[1-\kappa^{-1/2} tanh(\kappa^{1/2})]},
\end{equation}

\noindent for $\widetilde {\nu} = \nu$.

The drag ratio for porous spherical shells has been calculated by Qin and Kalony\cite{qin1993creeping} using the same approach.

Inspection of equations \ref{jones0} and \ref{Neale} show that the two models presented above yield different results. Both treatments yield a drag ratio $r(k)$ for the porous sphere, which decays with increasing permeability and tends to zero for very large permeabilities. For low permeability, $\kappa << 1$, the two analytical results, Eqs \ref{jones} and \ref{Neale}, coincide if Jones’s phenomenological parameter $\alpha$ is taken as equal to 1. In this case, both treatments yield:

\begin{equation}
r_{sphere,J-N}(k\ll 1) \approx [1-\kappa^{1/2}].
\end{equation}

\noindent
Curiously, $\alpha = 1$ is in accordance with the suggestion given in \cite{neale1974creeping}, on the basis of comparison with experiments.

However, for larger permeabilities ($\kappa \rightarrow 1$), the two approaches yield somewhat
different results, with Neale’s drag force decaying earlier in permeability with respect to Jones’s drag force. This can be seen in Fig. \ref{LBM-x-anal} of the next section.

\subsection{Lattice Boltzmann approach to porous flow}

In the LBM, fluid flow through porous objects is calculated in terms of the Reynolds number Re and porosity $\phi$. How do the probability densities $f_i (x, t)$ evolve as the fluid goes through porous objects? Dardis and McCloskey \cite{Dardis} proposed to assign a partial “bounce-back” step, in which the fluid particles are redirected according to the local solid fraction $n_s$, such that in the region defined as porous the $f_i$s are partially transmitted and partially reflected, with a reflection coefficient given by $n_s$. Thus, instead of a “Boolean” system of “solid” and “fluid” cells, a system of “porous” cells is used. Dardis and McCloskey \cite{Dardis}, and later Thorne and Sukop \cite{thorne2004lattice}, designed different proposals for the probabilistic bounce-back, which involved “in” and “out” flux, either from previous time-steps or from neighboring nodes. However, as demonstrated in Walsh, Burwinkle and Saar  \cite{walsh2009new}, the proposed scheme did not conserve mass and an adjustment was necessary. A simpler model that preserves the conservation of mass and which facilitates parallelization of codes was proposed \cite{walsh2009new}, in which the partial bounce-back step prior to the collision step is given by:

\begin{equation}
f_i(\mathbf{x},t)=n_sf_{-i}(\mathbf{x},t)+(1-n_s)f_i(\mathbf{x},t)
\end{equation}
where $-i$ indicates the opposite direction to $i$, and $n_s = 1 - \phi$ represents cell porosity. Thus, $n_s$ works as a transmission coefficient: if the cell at $x$ is fluid, $n_s = 0$, and normal propagation occurs; if the cell is solid, $n_s = 1$, and reflection occurs. Finally, if the cell belongs to a porous object, $ns \neq 1$, the fluid is partially reflected and partially propagated. In this approach, mass is conserved, as can be seen from the following equation:

\begin{eqnarray}
\rho &=& \sum f_i(\mathbf{x},t)=\sum_i n_sf_{i}(\mathbf{x},t)+\sum_i (1-n_s)f_i(\mathbf{x},t)\nonumber \\
&=&\rho n_s+ (1-n_s)\rho 
\end{eqnarray}

The drag force can be determined using equation \ref{FS}, with a label to account for momentum exchanges between fluid and obstacle. A matrix $\omega$ is introduced, with $\omega(x) = 0$ for a node occupied by fluid, and $\omega(x) = 1$ for a node occupied by the obstacle. Thus, equation \ref{FS}, which  becomes the force on a porous solid object, is given by \cite{Mei1}:

\begin{equation}\label{force-por}
 \mathcal{F}_s= \sum_{\mathbf{x}_s\in C_s}\sum_{i\neq 0}c_i(f_i(\mathbf{x}_b+\mathbf{c}_i\Delta t,t+\Delta t)+f_i(\mathbf{x}_b,t))(1-\omega(x_b)).
\end{equation}

It can be demonstrated \cite{Dardis, thorne2004lattice} that this model for the force on a porous solid object, using Eqs. \ref{densi} and \ref{force-por} reproduces Eq. \ref{darcy-velocity}, which corresponds to the flow in a porous medium
proposed by Johnson \cite{jones1973low}.
\section{LBM simulations: results for porous spheres and shells}

We performed LBM simulations for porous objects of spherical geometry for different Reynolds numbers ($Re <  1$). Data were obtained for the drag coefficient $C_d (k)$ and for the drag ratio $r(k)$ in terms of the dimensionless permeabilities $\kappa$.

\subsection {Definition of the simulations}

The spherical object with a radius $R$ is symmetrically arranged in a rectangular channel of dimensions $L \times H \times H$. Simulations require the definition and testing of a blockage ratio $B = R/H$ for which results become independent of size. Breuer et al. \cite{Breuer} use the blockage ratio $B = 1/6$ and $B = 1/8$ for square cylinders.

\begin{figure}[h]
\centering
\begin{quote}
\begin{center}
                \includegraphics[width=0.3\hsize]{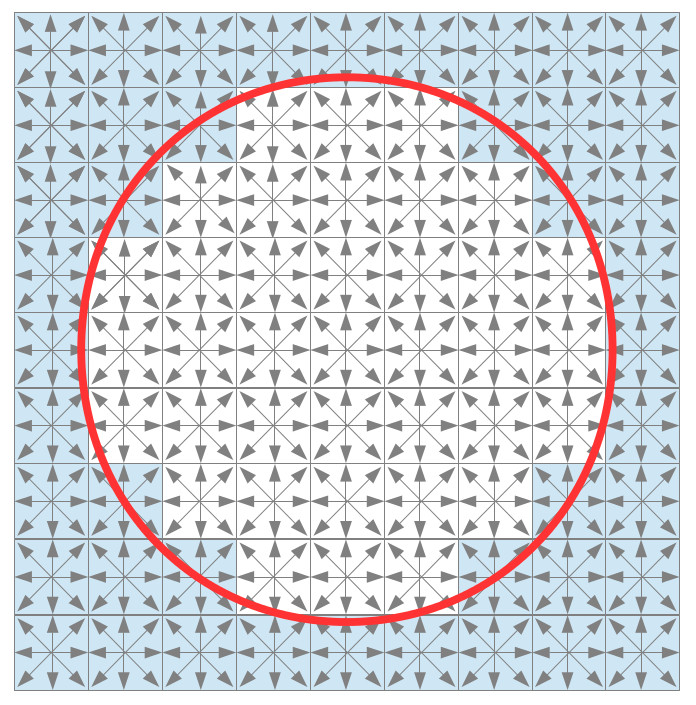} 
	   \caption{\textit{Porous sphere: $n_s=1-\phi$ for the white porous region,$n_s=0$ for the blue fluid region}}\label{figura:esf}
                  
\end{center}
\end{quote}
 \end{figure}

 \begin{figure}[h]
\centering
\begin{quote}
\begin{center}
                \includegraphics[width=0.3\hsize]{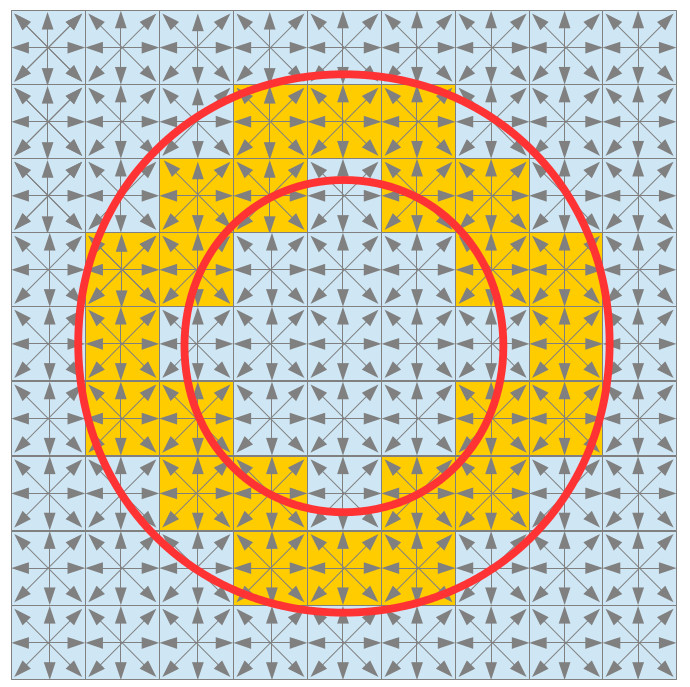} 
	   \caption{\textit{ LBM spherical shell. Blue denotes fluid cells and orange indicates porous cells.}}\label{figura:esf1}
                  
\end{center}
\end{quote}
 \end{figure}
 In our study, we take $B = 1/10$ for the spherical geometry, a value which has been tested by Mei et al. \cite{Mei1}. Toroidal boundary conditions are imposed on the upstream ($x = 0$) and downstream ($x = L$) fluxes \cite{Succi, sukop}, while sideways flux is “bounced-back”. The velocity of the fluid in the entry channel is defined by the Reynolds number, $Re \equiv \frac{U L}{\nu}$ . However, one must be careful to make $U\ll c_s$. Fluid viscosity is taken as fixed, with $\tau = 0.52$. 
 
Our algorithm and the resulting code were tested against well-known results for Poseuille flow, flow in a porous channel, diffusion as a function of time, and Stokes drag force on a solid sphere \cite{rodrigues2016coloides}.

As for the porous object, a value for $n_s$ is assigned to each node of the lattice in order to evaluate the propagation step in the porous regions. As shown in \cite{sukop}, a systematic error is present in the lattice approximation for the sphere, and interpolation techniques have been proposed to make the calculations more accurate\cite{mei1999accurate}. However, the error becomes negligible if the dimensionless drag ratio is used \cite{noymer1998drag}.

\subsection{Results and comparison to analytical predictions}

Here, we compare our results for the drag ratio $r(\kappa)$ (see definition Eq. \ref{r_k}) with different analytical predictions. We also examine the situation in which neither the particle radius $R$ nor its porosity $\Phi$ is known, as may be the case for several porous particle systems. In particular, we discuss the impact of our results on the calculation of particle radius from data and on its diffusion in the viscous fluid.

\begin{figure}[h]
\centering
\begin{quote}
\begin{center}
                \includegraphics[width=0.8\hsize]{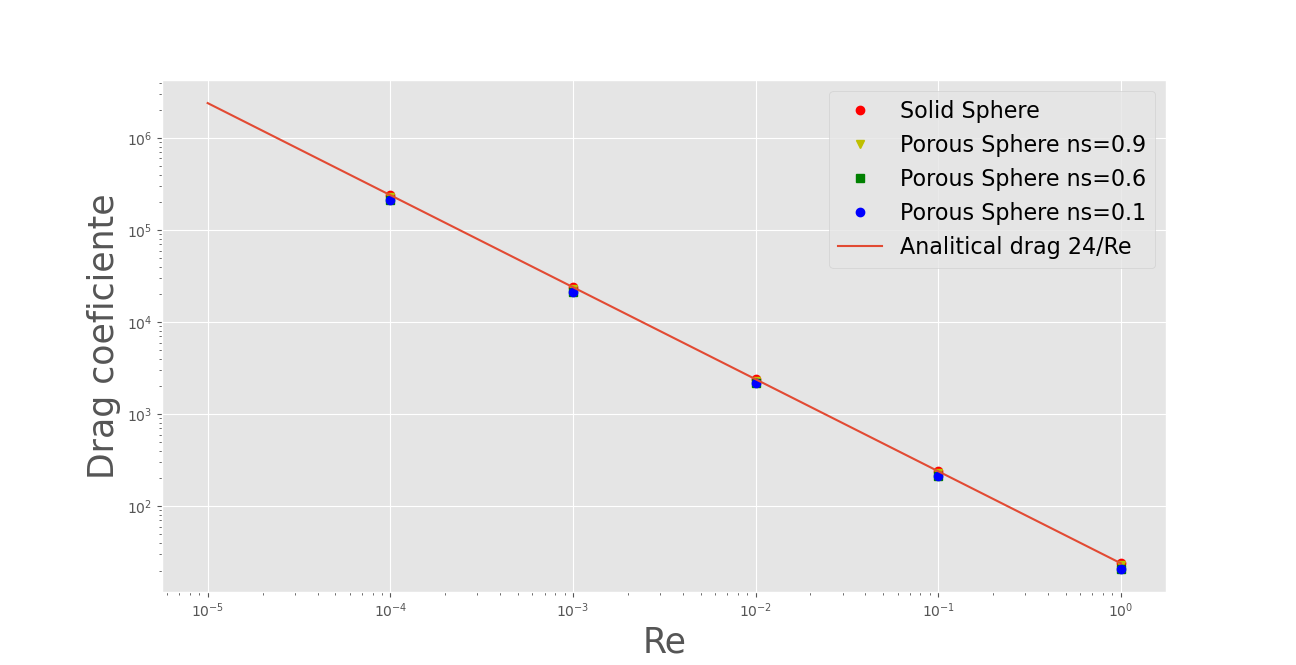} 
	   \caption{\textit{LBM drag coefficient x Reynolds number for porous spheres of different porosities. Full line gives analytical Stokes result for solid spheres Cd (k = 0) = 24/Re.}}\label{k-Re-poros}
                  
\end{center}
\end{quote}
 \end{figure}

Figure \ref{k-Re-poros} shows our data for the drag coefficient $Cd (k)$ (see Eq. \ref{Cd_k}) for a large range of Reynolds numbers ($Re \leq 1$) and for different porosities $\phi$ of the spherical obstacle. Our data faithfully reproduce the analytical results for impermeable spheres (Eq \ref{Cd_0}). It can be seen that the effect of porosity $\phi$ upon the drag coefficient $Cd (k)$ is to reduce it. However, there does not seem to be any effect of the Reynolds number, for $Re \leq 1$, as had already been seen in experiments for permeable spheres (see Fig. 5 of Masliyah et al. \cite{masliyah1980terminal} and Nandakumar et. al.\cite{nandakumar1982laminar}), as well as in previous computational results\cite{masliyah1980terminal} for permeable cylinders \cite{noymer1998drag, bhattacharyya2006fluid}.

In order to compare our results with existing analytical predictions, given in terms of reduced permeability $\kappa$ and presented in the previous section (Eqs. \ref{jones} \ref{Neale}), we need a relationship between permeability and porosity.  Porosity $\Phi$ and global permeability $k$ are related, but the relationship may depend on the morphology of the scattering objects. For LBM, Walsh and collaborators \cite{walsh2009new} have proposed the following equation:

\begin{equation}\label{k(Phi}
\bar{k}= \bar{\nu} \frac{\Phi }{2(1 - \Phi)}.
\end{equation}
noindent in which $\bar{k}$ and $\bar{\nu}$ are reduced permeability and viscosity, respectively.

 \begin{figure}[h]

\centering
\begin{quote}
\begin{center}
                \includegraphics[width=1.0\hsize]{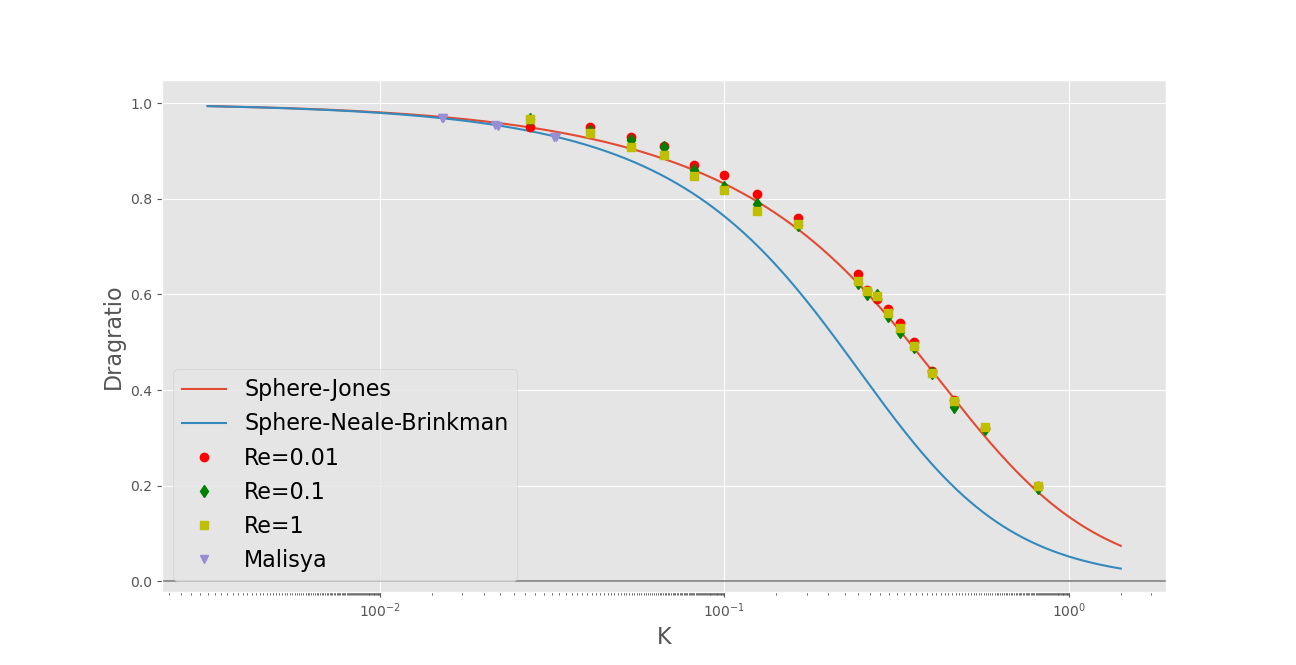} 
	   \caption{LBM drag ratio $\times$ permeability calculated for different Re numbers. Circles, diamonds and squares correspond to our LB results. Full line: Jones's anaytical prediction\cite{jones1973low} (Eq \ref{jones}). Dashed line: Neale's analytical prediction\cite{neale1974creeping} (Eq \ref{Neale}). Inverted triangles represent experimental results for intermediate porosities \cite{masliyah1980terminal}. }\label{LBM-x-anal}                 
\end{center}
\end{quote}
 \end{figure}
 
 Figure \ref{LBM-x-anal} shows our LBM results for the effect of permeability (in terms of the reduced permeability $\kappa$) on the drag ratio $r(\kappa)$ of a porous sphere, for Re in the range 0.0001 to 1.0. It can be seen that $r(\kappa)$ is very close to unity for $10^{-6}\leq \kappa \leq 10^{-4}$ , which means that the porous sphere is indistinguishable from a solid sphere. However, for larger permeabilities, $\kappa > 10^{-4} $, the drag ratio rapidly decreases.

A highly permeable sphere allows the fluid to pass with little resistance and the drag coefficient tends to zero as permeability increases. On the other hand, a sphere with low permeability allows little or no fluid to pass through.

Our data were compared with the analytical results of Jones \cite{jones1973low} and Neale and collaborators \cite{neale1974creeping} and with the experimental results of Masliyah \cite{masliyah1980terminal}. For low permeability, our numerical and the analytical results, as well as the experimental results are indistinguishable. However, for larger permeabilities ($\kappa > 10^{-3} $), the two analytical results diverge. Our results are in accordance with the results obtained in the analytical approach proposed by Jones \cite{jones1973low}, while large deviation is seen with respect to the prediction based on the Brinkmann model used by Neale and collaborators \cite{neale1974creeping}.

 \begin{figure}[h]
\centering
               \includegraphics[width=1.0\hsize]{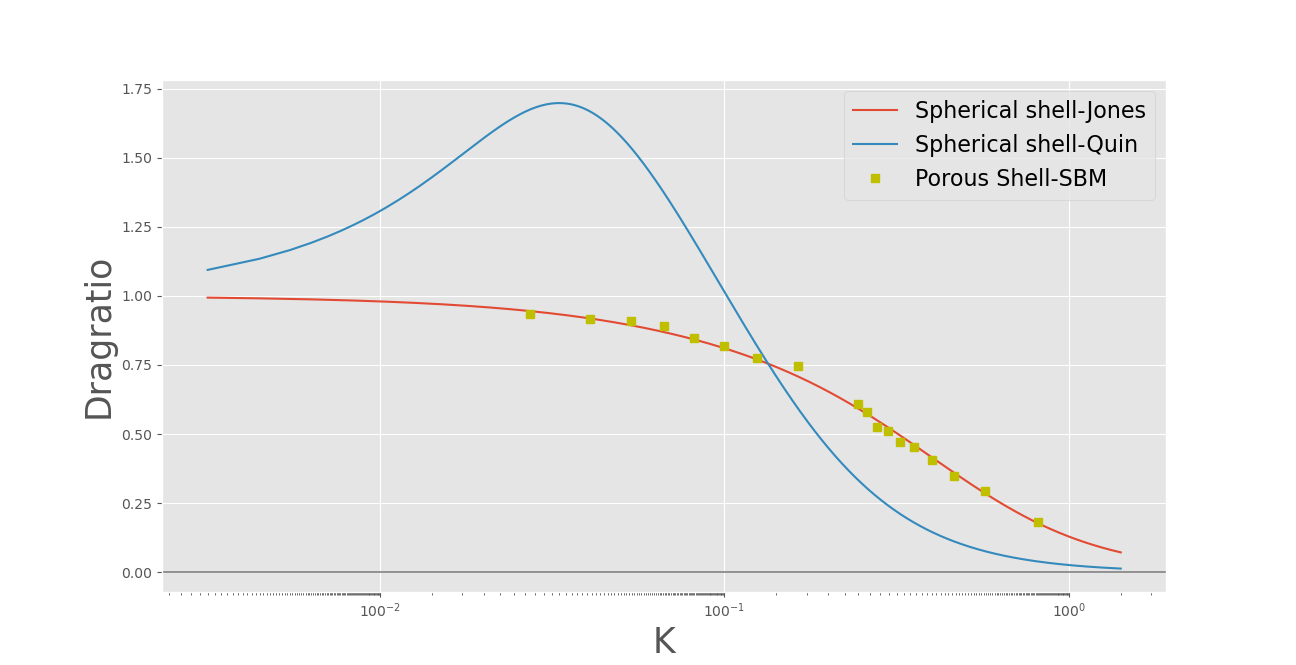} 
	     \caption{{\textit{LBM drag ratio r(k) $\times $ permeability for a porous
spherical shell (Re = 0.1, $\gamma = 0.2$). Full lines represents Jones's \cite{jones1973low} corresponding analytical solution (Eq \ref{jones0} ) and Quin's solution \cite{qin1993creeping} .}}} \label{shell}

 \end{figure}
 
 Figure \ref{shell} displays our LBM results for the drag ratio $r(k)$ of a porous spherical shell, for $Re=0.1$. Again, the simulation results are in good agreement with Jones’s\cite{jones1973low} analytical prediction. However, strong divergence is noted with respect to the results of Qin and collaborators \cite{qin1993creeping}, in which Brinkmann’s model was used for the porous fluid flow. As such, convergence with Jones’s analytical prediction is seen both for the sphere and the spherical shell.

 \begin{figure}[h]
\centering
         \includegraphics[width=1.0\hsize]{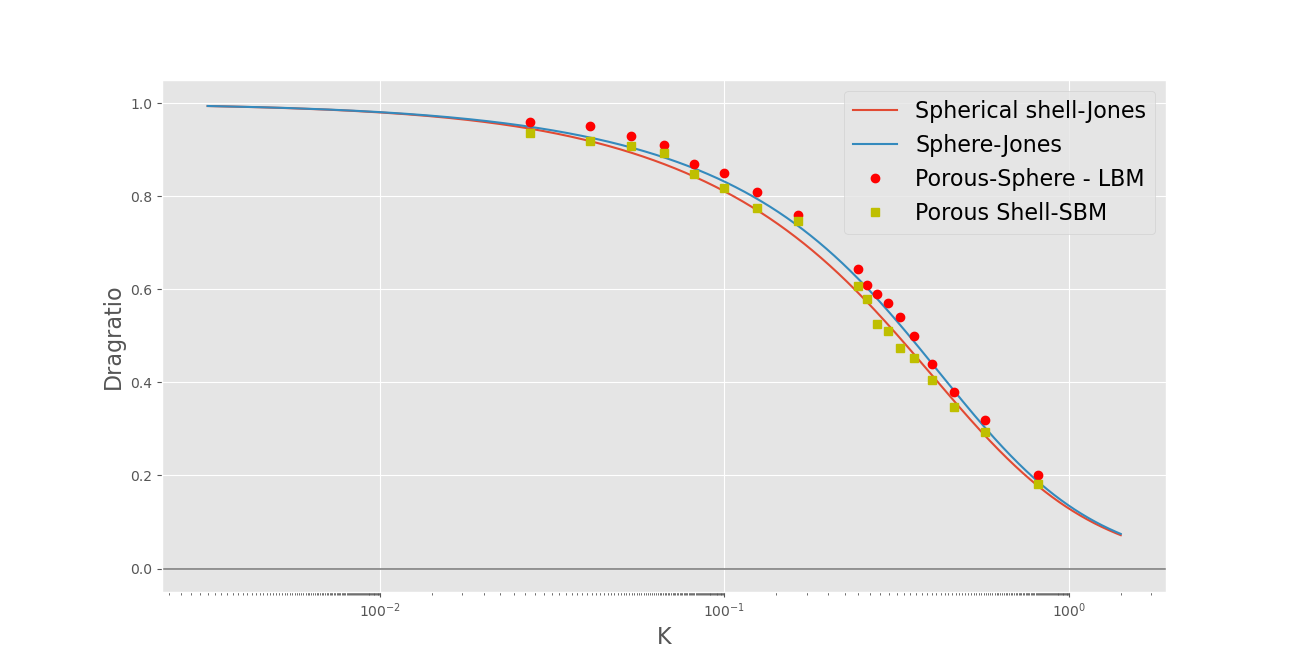} 
	  \caption{{\textit{LBM drag ratio r(k) $\times $ permeability for a porous sphere and for a porous spherical shell (Re = 0.1,$\gamma = 0.2$). Circles and squares indicate our LBM points. Full lines represents Jones's \cite{jones1973low} corresponding analytical solutions (Eqs \ref{jones0} and \ref{jones}). }}}\label{print1}

 \end{figure}
In Figure \ref{print1}, we show the drag ratio for both the porous sphere and the porous spherical shell obtained from our LBM simulations at $Re=0.1$. A small difference can be seen between the porous sphere and the porous spherical shell, with a slightly smaller drag ratio for the spherical shell. Jones’s analytical results  \cite{jones1973low} are also plotted (see equations \ref{jones0} and \ref{jones}), which present very good agreement with our results.

\section{Particle size and porosity from diffusion and drag for porous shells}
    Our study was initially inspired by the question of the determination of the radius of porous particles from DLS data. In this data the diffusion constant $D$ is obtained via light mirroring. The diffusion constant $D$ may be related to the drag force coefficient $b$, using Einstein’s relation, which is given by

\begin{equation}\label{Einstein}
D = \frac{k_B T}{b}.
\end{equation} 
For a solid sphere, the drag force calculated analytically by Stokes is well known\cite{landau2013fluid}. The linear viscous force coefficient is given by $b = 6\pi\mu R$; therefore, one may obtain the particle radius $R$ from:

\begin{equation} \label{Stokes radius}
 R = \frac{k_B T}{6 \pi \nu D}.
\end{equation}  

However, if the particle is porous, the Stokes relation is no longer valid, as we have seen above, and Eq. \ref{Stokes radius}, frequently used for the interpretation of DLS data (see Eq. 17 in Chen et al. \cite{chen2012characterizing}), will yield false results on the particle dimension $R$. Thus, one may need some form of calculation of fluid flow through pores in order to correctly interpret the diffusion data in terms of the nanoparticle size.

In the analytical treatment of flow through porous objects, permeability and particle size are entangled in the reduced variable $\kappa \equiv \frac{k}{R^2}$, as can be seen in Eqs. \ref{jones0} and \ref{Neale}. Thus, complete description depends on knowledge of both particle size $R$ and permeability $k$. However, permeability $k$ is not easily established from experiments.

If porous nanoparticles are put to technological use, particle structure must be investigated. Radius and porosity are two of the characteristic features of such nanoparticles. 

\begin{figure}[h]
\centering
               \includegraphics[width=1.0\hsize]{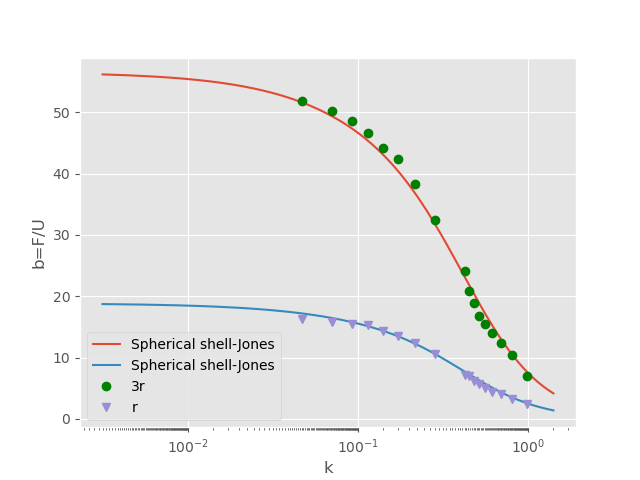} 
	     \caption{\textit{Drag force vs reduced permeability $\kappa$ for different particle radii. Lines represent Jones's result (Eqs. \ref{jones0} and \ref{jones}) and figures represent our LBM data (triangles for radius $r$ and spheres for radius $3r$).}}
\label{b-x-k}
 \end{figure}

In order to examine the effect of porosity on particle radius $R$, we compared the drag force coefficient $b$ for spheres of different radii. As can be seen in Fig. \ref{b-x-k}, there is an overlap of values for the viscous force $\mathcal{F}_d$ for different permeabilities and radii.

\begin{figure}[h]
\centering
               \includegraphics[width=0.7\hsize]{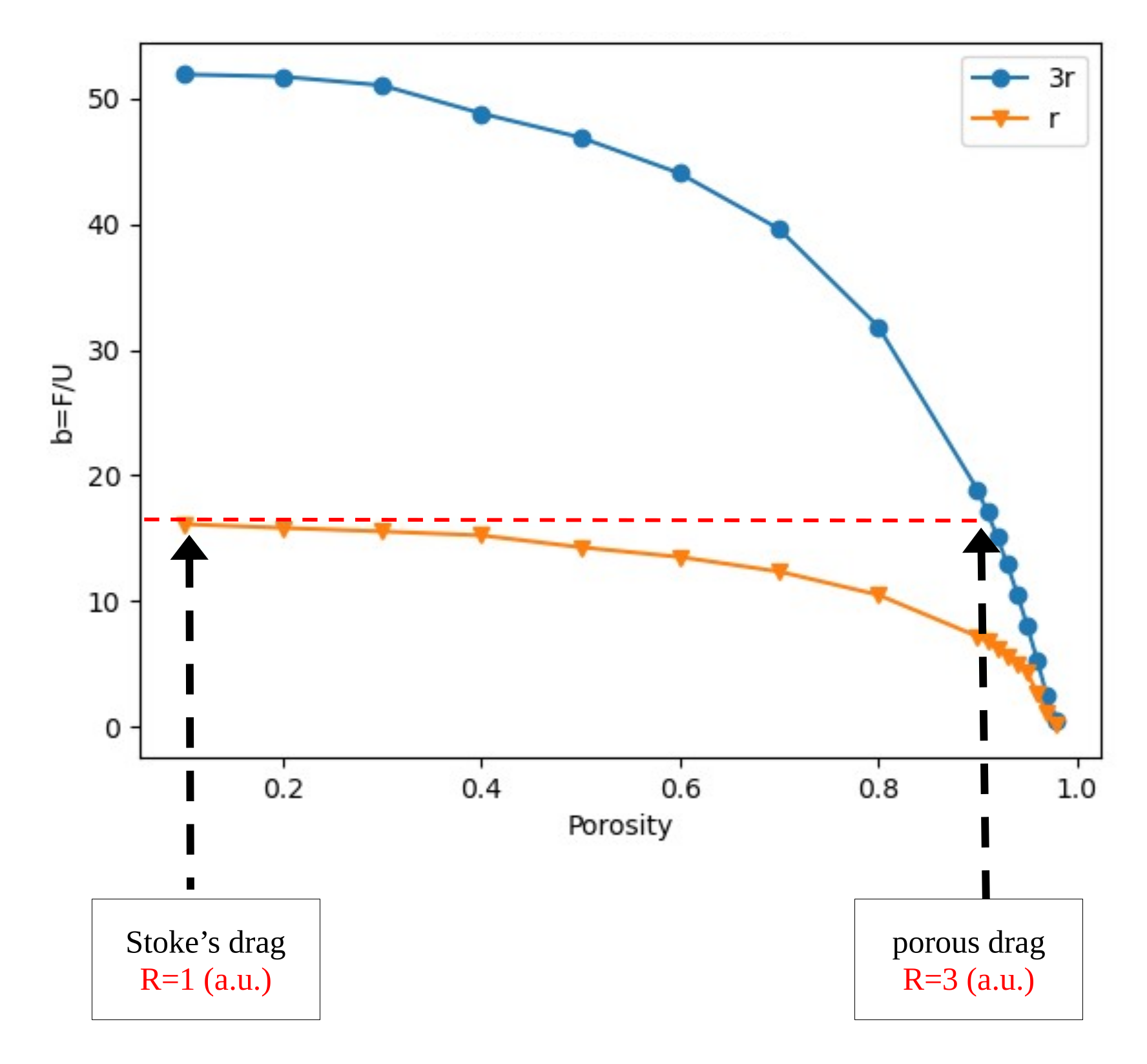} 
	     \caption{{\textit{Drag force vs porosity.}}}\label{FvsP}

 \end{figure}

In Fig. \ref{FvsP}, we emphasize the importance of a thorough investigation of porosity in order to obtain particle radius from drag force. We compare the drag forces $\mathcal{F}_d$ on particles of radius $1$ and $3$ and show that they may present the same friction coefficient $b$ for different porosities $\Phi$. In particular, a zero-porosity particle of a radius $1$ and a $90\%$ porous particle of a radius $3$ will present the same value as $b$, approximately $18$ (in arbitrary units).

This is an interesting result for comparison to the data of Enoki et al.\cite{Thais} on the radius of a model dissociating membrane in a solution at different temperatures. The lipid vesicle radius was investigated through SLS) and DLS. Data from SLS indicated a range of temperatures at which the vesicles would swell, with the radius increasing from $30nm$ to nearly $90nm$, while DLS indicated that the radius would gain a much smaller increase, of nearly $10 \%$ over $30nm$. In SLS, the diamenter is obtained directly from the plot of the scattering data \cite{Zimm}. As for DLS, as discussed above, radius dimension is inferred from the diffusion constant $D$ and the Stokes-Einstein relation (Eq. \ref{Stokes radius}). The authors of the paper considered the possibility of a lower drag for a larger membraned shell due to porosity. Adopting the $90nm$ radius obtained from the SLS data as correct, Enoki et al. used mass conservation to calculate the amount of porosity and reached a value of $90 \%$. Fig. \ref{FvsP} shows that a solid sphere of a radius $1$ and a $90\%$ porous sphere of a radius $3$ present the same drag coefficient and this lends support to the rationale of the authors on the discrepancy between SLS and DLS data on the lipid structure radius.

  \section{Discussion and conclusions}

 Herein, we present a Lattice Boltzmann approach for the study of porous particle sizes as an alternative to the macroscopic approach based on the Navier-Stokes hydrodynamic equation, which also requires numerical procedures.

 The hydrodynamic treatment of the drag force on porous spheres was the subject of previous analytical studies, but different choices yield different results \cite{jones1973low,brinkman1949calculation, neale1974creeping} in the case of higher porosities: the divergence needs to be cleared up.  We were able to show that the discretized Boltzmann approach may serve  to check this question. For the range of Reynolds numbers, we considered ($Re<1$), and our results for the drag force present a good fit to the results of the analytical solution for the hydrodynamic equations, in which flow through the porous object is considered in terms of the Darcy equation (\ref{darcy}) under generalization of semi-empirical boundary conditions proposed by Beavers and Joseph \cite{Beavers}, and used by Jones \cite{jones1973low}. If compared to our simulation results, the alternative analytical solution of the Brinkman equation \cite{brinkman1949calculation} for the fluid in the porous medium, given by Neale and collaborators \cite{neale1974creeping}, underestimates the drag force suffered by porous spheres with high porosities. Thus, our results support Jones’s proposal (Eq. \ref{jones}) for the drag ratio of porous spherical particles.

In both of the analytical solutions, the drag force is obtained as a function of a dimensionless permeability, which is given by the ratio of permeability to the square of the particle radius. In the case of our LBM calculations, it is simple to disentangle the two dependencies of the drag force. Our data demonstrate that the same value for the drag force may lead to different values for the particle radius $R$, depending on particle porosity. This result implies that care must be taken in the interpretation of the relationship between drag force obtained from experimental data and the true particle size.  If the expression for the drag force on a solid sphere is used (Stokes equation), one might inappropriately obtain a smaller effective diameter for the porous particle. In particular, special care must be taken when using DLS for particle size determination. The presence of porosity must be checked using other techniques, and only then may DLS data be interpreted.

A test of our proposal and a byproduct of our investigation was the checking of the hypothesis on the degree of pore formation of DMPG liposomes, in the order-disorder transition temperature region of the vesicle solution. Enoki and collaborators \cite{Thais} compared results for particle size from DLS and SLS. The first technique yielded a radius for the DMPG liposome almost three times smaller than the radius obtained using the second technique. The divergence was interpreted by the authors in terms of pore formation. Particle size may be obtained directly from SLS data, through Zimm plots; whereas, in the case of DLS, particle size must be deduced from data on the particle diffusion constant. Taking the SLS radius as the correct value (near $30 nm$), the authors calculated a value for porosity close to $90\%$, considering the preservation of lipid volume. Our numerical measurements for two spherical shells with radii $r$ and $3r$, shown in Fig. \ref{FvsP}, clearly lend support to the hypothesis of the $90\%$ liposome porosity. Our results strongly suggest a possible source of error that has generated inconsistencies in DLS measurements when compared with other techniques.  

	In conclusion, porous particle size measurements using DLS require a second technique for the determination of the degree of porosity, as well as a fluid flow calculator for the relationship between the drag force and the particle radius. We suggest that an LBM code may be a good “calculator” to be coupled to the DLS technique for the interpretation of data, since other studies also suggest that DLS may underestimate the capsule hydrodynamic radius \cite{chen2012-pore-structure-various-techniques}.

\section*{Acknowledgement}
The authors  thanks support from the Physics Institute of the University of São Paulo and Federal University of Amazonas. We thanks Dr. T. Enoki for enlightening discussions experimental techniques for lipid membranes.


\begin{thebibliography}{56}
\bibitem{jones1973low}
Jones, I. P., \textit{Low Reynolds number flow past a porous spherical shell}, Mathematical Proceedings of the Cambridge Philosophical Society, vol. 73, no. 01, pp. 231--238, 1973, Cambridge Univ Press.

\bibitem{jaiswal2015stokes}
Jaiswal, Bharat Raj and Gupta, Bali Ram, \textit{Stokes flow over composite sphere: Liquid core with permeable shell}, Journal of Applied Fluid Mechanics, vol. 8, no. 3, pp. 339--350, 2015.

\bibitem{hewitt2016flow}
Hewitt, Duncan R., Nijjer, Japinder S., Worster, M. Grae, and Neufeld, Jerome A., \textit{Flow-induced compaction of a deformable porous medium}, Physical Review E, vol. 93, no. 2, p. 023116, 2016.

\bibitem{Thais}
Enoki, Thais A., Henriques, Vera B., and Lamy, M. Teresa, \textit{Light scattering on the structural characterization of DMPG vesicles along the bilayer anomalous phase transition}, Chemistry and Physics of Lipids, vol. 165, no. 8, pp. 826--837, 2012.

\bibitem{Barroso}
Barroso, Rafael P., Riske, Karin A., Henriques, Vera B., and Lamy, M. Teresa, \textit{Ionization and Structural Changes of the DMPG Vesicle along Its Anomalous Gel Fluid Phase Transition: A Study with Different Lipid Concentrations}, Langmuir, vol. 26, pp. 13805--13814, 2010.

\bibitem{HF}
Horbach, J. and Frenkel, D., \textit{Lattice Boltzmann Method for simulation of transport phenomena in charged colloids}, Phys. Rev. E, vol. 64, 2001.

\bibitem{horbach}
Horbach, J{\"u}rgen and Frenkel, Daan, \textit{Lattice-Boltzmann method for the simulation of transport phenomena in charged colloids}, Physical Review E, vol. 64, no. 6, pp. 061507, 2001.

\bibitem{zhang2008red}
Zhang, Junfeng, Johnson, Paul C., and Popel, Aleksander S., \textit{Red blood cell aggregation and dissociation in shear flows simulated by lattice Boltzmann method}, Journal of Biomechanics, vol. 41, no. 1, pp. 47--55, 2008.

\bibitem{krafczyk1998analysis}
Krafczyk, M., Cerrolaza, M., Schulz, M., and Rank, E., \textit{Analysis of 3D transient blood flow passing through an artificial aortic valve by Lattice--Boltzmann methods}, Journal of Biomechanics, vol. 31, no. 5, pp. 453--462, 1998.

\bibitem{riske2009extensive}
Riske, Karin A., Amaral, Lia Q., and Lamy, M. Teresa, \textit{Extensive bilayer perforation coupled with the phase transition region of an anionic phospholipid}, Langmuir, vol. 25, no. 17, pp. 10083--10091, 2009.

\bibitem{chen2012characterizing}
Chen, Zhi Hong, Kim, Chanhoi, Zeng, Xiang-bing, Hwang, Sun Hye, Jang, Jyongsik, and Ungar, Goran, \textit{Characterizing size and porosity of hollow nanoparticles: SAXS, SANS, TEM, DLS, and adsorption isotherms compared}, Langmuir, vol. 28, no. 43, pp. 15350--15361, 2012.

\bibitem{kaasalainen2017size}
Kaasalainen, Martti, Aseyev, Vladimir, von Haartman, Eva, Karaman, Didem {\c{S}}en, M{\"a}kil{\"a}, Ermei, Tenhu, Heikki, Rosenholm, Jessica, and Salonen, Jarno, \textit{Size, stability, and porosity of mesoporous nanoparticles characterized with light scattering}, Nanoscale Research Letters, vol. 12, pp. 1--10, 2017.

\bibitem{karin}
Riske, Karin A., Barroso, Rafael P., Vequi-Suplicy, C{\^a}ntia C., Germano, Renato, Henriques, Vera B., and Lamy, M. Teresa, \textit{Lipid bilayer pre-transition as the beginning of the melting process}, Biochemica et Biophysica Acta. Biomembranes, vol. 1788, pp. 954--963, 2009.

\bibitem{neale1974creeping}
Neale, G., Epstein, N., and Nader, W., \textit{Creeping flow relative to permeable spheres}, Chemical Engineering Science, vol. 29, no. 5, p. 1352, 1974.

\bibitem{kerson1963statistical}
Kerson, Huang, \textit{Statistical mechanics}, Wiley Eastern Limited, 1963.

\bibitem{Chen}
Doolen, S. Chen and G. D., \textit{Lattice Boltzmann method for fluid flows}, Annu. Rev. Fluid Mech., pp. 329--364, 1998.

\bibitem{Higuera}
Higuera, F. J., Succi, S., and Benzi, R., \textit{Lattice gas dynamics with enhanced collisions}, Europhys. Lett., vol. 9, p. 345, 1989.

\bibitem{Wolf}
Wolf-Gladrow, D. A., \textit{Lattice-Gas Cellular Automata and Lattice Boltzmann Models an Introduction}, Springer, Germany, 2000.

\bibitem{ziegler1993boundary}
Ziegler, Donald P., \textit{Boundary conditions for lattice Boltzmann simulations}, Journal of Statistical Physics, vol. 71, no. 5-6, pp. 1171--1177, 1993.

\bibitem{he}
He, X. and Luo, L.S., \textit{On the theory of the lattice Boltzmann equation: From the Boltzmann equation to the lattice Boltzmann equation}, Phys. Rev. E, vol. 56, p. 6811, 1997.

\bibitem{mei1999accurate}
Mei, Renwei, Luo, Li-Shi, and Shyy, Wei, \textit{An accurate curved boundary treatment in the lattice Boltzmann method}, Journal of Computational Physics, vol. 155, no. 2, pp. 307--330, 1999.

\bibitem{filippova1998grid}
Filippova, Olga, and H{\"a}nel, Dieter, \textit{Grid refinement for lattice-BGK models}, Journal of Computational Physics, vol. 147, no. 1, pp. 219--228, 1998.

\bibitem{inamuro1995non}
Inamuro, Takaji, Yoshino, Masato, and Ogino, Fumimaru, \textit{A non-slip boundary condition for lattice Boltzmann simulations}, Physics of Fluids (1994-present), vol. 7, no. 12, pp. 2928--2930, 1995.

\bibitem{hassan}
Farhat, Hassan; Lee, Joon Sang; Kondaraju, Sasidhar, \textit{Accelerated Lattice Boltzmann Model for Colloidal Suspensions}, Springer, 2014.

\bibitem{monteferrante2014lattice}
Monteferrante, Michele, Melchionna, Simone, and Marconi, Umberto Marini Bettolo, \textit{Lattice Boltzmann method for mixtures at variable Schmidt number}, The Journal of Chemical Physics, vol. 141, no. 1, p. 014102, 2014.

\bibitem{yoshida2014transmission}
Yoshida, Hiroaki and Hayashi, Hidemitsu, \textit{Transmission--reflection coefficient in the lattice Boltzmann method}, Journal of Statistical Physics, vol. 155, no. 2, pp. 277--299, 2014.

\bibitem{prestininzi2015reassessing}
Prestininzi, Pietro, Montessori, Andrea, La Rocca, Michele, and Succi, Sauro, \textit{Reassessing the single relaxation time Lattice Boltzmann Method for the simulation of Darcy's flows}, International Journal of Modern Physics C, p. 1650037, 2015.

\bibitem{walsh2009new}
Walsh, Stuart DC, Burwinkle, Holly, and Saar, Martin O, \textit{A new partial-bounceback lattice-Boltzmann method for fluid flow through heterogeneous media}, Computers and Geosciences, vol. 35, no. 6, pp. 1186--1193, 2009.

\bibitem{masliyah1980terminal}
Masliyah, Jacob H and Polikar, Marcel, \textit{Terminal velocity of porous spheres}, The Canadian Journal of Chemical Engineering, vol. 58, no. 3, pp. 299--302, 1980, Wiley Online Library.

\bibitem{Beavers}
Beavers, Gordon S and Joseph, Daniel D, \textit{Boundary conditions at a naturally permeable wall}, Journal of Fluid Mechanics, vol. 30, no. 01, pp. 197--207, 1967, Cambridge Univ Press.

\bibitem{he1997analytic}
He, Xiaoyi and Zou, Qisu and Luo, Li-Shi and Dembo, Micah, \textit{Analytic solutions of simple flows and analysis of nonslip boundary conditions for the lattice Boltzmann BGK model}, Journal of Statistical Physics, vol. 87, no. 1-2, pp. 115--136, 1997, Springer.

\bibitem{Cercignani1}
Cercignani, Carlo, \textit{Mathematical Methods in Kinetic Theory}, Plenum Press, 1969.

\bibitem{Qian}
Qian, Y. H., D'Humi{\`e}res, D., and Lallemand, P., \textit{Lattice BGK Models for Navier-Stokes Equation}, Europhys Lett., vol. 1788, no. 6, p. 479, 1991.

\bibitem{mei}
Mei, Renwei and Shyy, Wei. and Yu, Dazhi and Luo, Li-Shi, \textit{Lattice Boltzmann Method for 3D Flows with Curved Boundary}, Journal of Computational Physics, vol. 161, no. 2, pp. 680--699, 2000.

\bibitem{Succi}
Succi, S., \textit{The Lattice Boltzmann Equation for Fluid Dynamics and Beyond}, Claderon Press Oxford, 2001.

\bibitem{landau2013fluid}
Landau, Lev Davidovich and Lifshitz, Evgenii Mikhailovich, \textit{Fluid Mechanics: Landau and Lifshitz: Course of Theoretical Physics, Volume 6}, Elsevier, 2013.

\bibitem{qin1993creeping}
Qin, Yu and Kaloni, PN, \textit{Creeping flow past a porous spherical shell}, ZAMM-Journal of Applied Mathematics and Mechanics/Zeitschrift für Angewandte Mathematik und Mechanik, vol. 73, no. 2, pp. 77--84, 1993, Wiley Online Library.

\bibitem{Dardis}
Dardis, Orla and McCloskey, John, \textit{Lattice Boltzmann scheme with real numbered solid density for the simulation of flow in porous media}, Phys. Rev. E, vol. 57, no. 4, p. 4834, Apr. 1998.

\bibitem{thorne2004lattice}
Thorne, DT and Sukop, MC, \textit{Lattice Boltzmann model for the Elder problem}, Developments in Water Science, vol. 55, pp. 1549--1557, 2004, Elsevier.

\bibitem{Ladd}
Ladd, A. J. C., \textit{Short-time motion of colloidal particles: Numerical simulation via a fluctuating lattice-Boltzmann equation}, Physical Review Letters, vol. 70, no. 9, p. 1339, 1993.

\bibitem{Mei1}
Mei, R., Yu, D., and Shyy, W., \textit{Force evaluation in the Lattice Boltzmann method involving curved geometry}, Phys. Rev. E, vol. 65, Apr. 2002.

\bibitem{sukop}
Sukop, M. C. and {Thorne Jr.}, D. T, \textit{Lattice Boltzmann Modeling: An Introduction for Geoscientists and Engineers}, Springer, Berlin, 2006.
\bibitem{rodrigues2016coloides}
Rodrigues Junior, Wagner Gomes, \textit{Coloides carregados ou porosos: estudos das propriedades hidrodinâmicas e eletrocinéticas com o método Lattice Boltzmann}, PhD thesis, Universidade de São Paulo, 2016.

\bibitem{noymer1998drag}
Noymer, Peter D and Glicksman, Leon R and Devendran, Anand, \textit{Drag on a permeable cylinder in steady flow at moderate Reynolds numbers}, Chemical Engineering Science, vol. 53, no. 16, pp. 2859--2869, 1998, Elsevier.

\bibitem{bhattacharyya2006fluid}
Bhattacharyya, S and Dhinakaran, S and Khalili, Arzhang, \textit{Fluid motion around and through a porous cylinder}, Chemical Engineering Science, vol. 61, no. 13, pp. 4451--4461, 2006, Elsevier.

\bibitem{nandakumar1982laminar}
Nandakumar, K and Masliyah, Jacob H, \textit{Laminar flow past a permeable sphere}, The Canadian Journal of Chemical Engineering, vol. 60, no. 2, pp. 202--211, 1982, Wiley Online Library.

\bibitem{brinkman1949calculation}
Brinkman, Hendrik C, \textit{A calculation of the viscous force exerted by a flowing fluid on a dense swarm of particles}, Flow, Turbulence and Combustion, vol. 1, no. 1, pp. 27--34, 1949, Springer.

\bibitem{Breuer}
Breuer, M., Bernsdorf, J., Zeiser, T., and Durst, F., \textit{Accurate computations of the laminar flow past a square cylinder based on two different methods: lattice-Boltzmann and finite-volume}, International Journal of Heat and Fluid Flow, vol. 21, pp. 186--196, 2000.

\bibitem{Batchelor}
Batchelor, G.K., \textit{Introduction to Fluid Dynamics}, Cambridge University Press, 2000.

\bibitem{baeza2017recent}
Baeza, Alejandro and Ruiz-Molina, Daniel and Vallet-Regí, María, \textit{Recent advances in porous nanoparticles for drug delivery in antitumoral applications: inorganic nanoparticles and nanoscale metal-organic frameworks}, Expert Opinion on Drug Delivery, vol. 14, no. 6, pp. 783--796, 2017, Taylor \& Francis.

\bibitem{Sayed2017-porous-particle-review}
Sayed, E., Haj-Ahmad, R., Ruparelia, K., et al., \textit{Porous Inorganic Drug Delivery Systems—a Review}, PharmSciTech, vol. 18, pp. 1507--1525, 2017, AAPS Publications.

\bibitem{bear2012-livro}
Bear, Jacob and Bachmat, Yehuda, \textit{Introduction to Modeling of Transport Phenomena in Porous Media}, vol. 4, Springer Science \& Business Media, 2012.

\bibitem{vafai2015handbook}
Vafai, Kambiz, \textit{Handbook of Porous Media}, CRC Press, 2015.

\bibitem{chen2012-pore-structure-various-techniques}
Chen, Zhi Hong and Kim, Chanhoi and Zeng, Xiang-bing and Hwang, Sun Hye and Jang, Jyongsik and Ungar, Goran, \textit{Characterizing size and porosity of hollow nanoparticles: SAXS, SANS, TEM, DLS, and adsorption isotherms compared}, Langmuir, vol. 28, no. 43, pp. 15350--15361, 2012, ACS Publications.

\bibitem{kaasalainen2017-structure-lightscatt}
Kaasalainen, Martti and Aseyev, Vladimir and von Haartman, Eva and Karaman, Didem Şen and Mäkilä, Ermei and Tenhu, Heikki and Rosenholm, Jessica and Salonen, Jarno, \textit{Size, stability, and porosity of mesoporous nanoparticles characterized with light scattering}, Nanoscale Research Letters, vol. 12, pp. 1--10, 2017, Springer.
\bibitem{guo2002LB}
Guo, Zhaoli and Zhao, TS, \textit{Lattice Boltzmann model for incompressible flows through porous media}, Physical Review E, vol. 66, no. 3, p. 036304, 2002, APS.

\bibitem{heimburg-book}
Heimburg, Thomas, \textit{Thermal Biophysics of Membranes}, Wiley, 2007.

\bibitem{huang2015multiphase}
Huang, Haibo and Sukop, Michael and Lu, Xiyun, \textit{Multiphase lattice Boltzmann methods: Theory and application}, John Wiley \& Sons, 2015.

\bibitem{zhang2011lattice}
Zhang, Junfeng, \textit{Lattice Boltzmann method for microfluidics: models and applications}, Microfluidics and Nanofluidics, vol. 10, pp. 1--28, 2011, Springer.

\bibitem{lamy2003peculiar}
Lamy-Freund, M Teresa and Riske, Karin A, \textit{The peculiar thermo-structural behavior of the anionic lipid DMPG}, Chemistry and Physics of Lipids, vol. 122, no. 1-2, pp. 19--32, 2003, Elsevier.

\bibitem{Zimm}
Zimm, B.H., \textit{Molecular Theory of the Scattering of Light in Fluids}, J. Chem. Phys., vol. 13, no. 4, pp. 141--145, 1945.
\end{thebibliography}
\end{document}